\documentclass[12pt]{article}

\usepackage{newtxtext,newtxmath}
\usepackage{amsmath}

\usepackage{graphicx}

\usepackage[letterpaper,margin=1in]{geometry}

\renewenvironment{abstract}
	{\quotation}
	{\endquotation}

\date{}

\makeatletter
\renewcommand{\fnum@figure}{\textbf{Figure \thefigure}}
\renewcommand{\fnum@table}{\textbf{Table \thetable}}
\makeatother

\usepackage{scicite}

\usepackage{url}

\usepackage[utf8]{inputenc}
\usepackage{mathtools}

\def\scititle{
	Direct identification of local doping effects in Barium-hexaferrite by electron vortex beams
}
\title{\bfseries \boldmath \scititle}

\author{
	Darius Pohl$^{1\ast\dagger}$,\and
	Hitoshi Makino$^{1}$,\and
	Arthur Ernst$^{2,3,4}$,\and
	Devendra Singh Negi$^{5}$,\and
	Sebastian Schneider$^{1}$,\and
	Rolf Erni$^{6}$,\and
	Jan Rusz$^{7}$ \\
	\small$^{1}$DCN, TUD Dresden University of Technology, Dresden, Germany.\and
	\small$^{2}$Johannes Kepler University, Linz, Austria.\and
    \small$^{3}$ Max-Planck-Institut für Mikrostrukturphysik, Halle (Saale), Germany.\and
    \small$^{4}$ Donostia International Physics Center (DIPC), Donostia-San Sebasti\'{a}n, Spain.\and
	\small$^{5}$ Department of Materials Engineering, Indian Institute of Technology, Jodhpur, India.\and
        \small$^{6}$Electron Microscopy Center, Empa, \\ \small Swiss Federal Laboratories for Materials Science and Technology, Dübendorf, Switzerland.\and
	  \small$^{7}$Department of Physics and Astronomy, Uppsala University, Uppsala, Sweden.\and
	\small$^\ast$Corresponding author. Email: darius.pohl@tu-dresden.de\and
}

\begin{document} 

\maketitle

\begin{abstract} \bfseries \boldmath
We demonstrate atomic-scale mapping of local magnetic moments and doping effects in Ti-doped barium hexaferrite ($\rm BaFe_{11}TiO_{19}$) using atom-sized electron vortex beams (EVBs) with controlled orbital angular momentum (OAM) in a scanning transmission electron microscope. By measuring electron energy loss magnetic circular dichroism (EMCD) at the $\rm Fe-L_{2,3}$ edges, we directly resolve the spatial distribution of antiparallel-aligned magnetic sublattices and quantify the impact of non-magnetic $\rm Ti^{4+}$ substitution. The EMCD signal, detected from single atomic Fe columns, reveals a marked reduction and sign reversal in the magnetic moment at Ti-rich $\rm 4f_2$ sites, corroborated by inelastic scattering simulations and density functional theory calculations that indicate induced $\rm Fe^{2+}$ formation and modified exchange interactions. Our results show that EVBs enable direct, element-specific, and atomically resolved magnetic characterization, opening new avenues for investigating local magnetic phenomena and dopant effects in nano-structured magnetic materials, such as those used in spintronic devices. This method paves the way for detailed studies of complex spin textures, magnetic interfaces, and dynamic processes at the atomic scale.
\end{abstract}

\noindent
Magnetism has fascinated scientists for centuries, driving innovations from the compass to modern data storage. Yet, the nanoscale particularities of magnetic materials like finite-size, surface effects \cite{batlle2002} and curved nanomagnetism \cite{fernandez2017} remain a frontier of research, with implications for emerging technologies like spintronics \cite{sinova2012} and quantum computing. At the heart of this exploration lies the challenge of probing magnetic properties with atomic precision. 
Electron vortex beams (EVBs) are promising candidates to reach that goal in a transmission electron microscope \cite{bliokh2007,uchida2010,verbeeck2010,mcmorran2011}. EVBs possess an orbital angular momentum (OAM) $\boldsymbol\ell$ which stems from the azimuthally increasing phase of the electron wave. Common methods for the generation of EVBs are holographic diffraction masks (absorption and phase masks), spiraling phase plates, (switchable) magnetic needles, or aberration correctors \cite{verbeeck2014,beche2014,Clark2013,idrobo2016}. In all cases, the beam consists of a superposition of contributions with differently weighted OAM, which defines their OAM purity \cite{schachinger2021}. Fig. 1A shows how the path of rays in the electron microscope is manipulated in order to produce isolated EVBs with user-defined OAM and high purity \cite{pohl2017}. The interaction of an EVB with the spin system of a magnetic sample is described by the double differential scattering cross-section (DDSCS), which in the case of EVBs and their specific phase symmetry enhances the magnetic scattering contribution that can be measured by a method called electron energy loss magnetic circular dichroism (EMCD) \cite{schattschneider2006,hebert2003,Rusz2013,Rusz2014,schattschneider2014}.  The imbalance in the density of states (DOS) of the spin-up and spin-down channels of ferrimagnets and ferromagnets, gives rise to dichroism in the electron energy loss spectra (EELS) of EVBs with different OAM or of magnetic moments pointing in different directions. The element specific EMCD signal is expected to be of the order of 5\% for a ferromagnetic system \cite{Rusz2014,pohl2015}. \\
In this study we used EVBs to detect the EMCD of single atomic Fe columns in the Ti doped barium hexaferrite, $\rm BaFe_{11}TiO_{19}$ – a ferrimagnetic material with a large magneto crystalline anisotropy and a high Curie temperature which is of great interest for magnetic recording, microwave magnetic devices, and permanent magnets \cite{pullar2012}. In comparison with inelastic scattering simulation and density functional theory (DFT) calculations the effect of non-magnetic $\rm Ti^{4+}$ doping on the magnetic ordering is resolved.

\subsection*{Electron Vortex Beam Spectroscopy}
To detect an EMCD signal using EVBs in an electron microscope, we choose the well-studied ferrimagnet $\rm BaFe_{11}TiO_{19}$ \cite{brabers1999}. The large lattice parameter of the ferrimagnets (space group P63/mmc, a = 5.929~$\Angstrom$, c = 23.413~$\Angstrom$), 
their native resistance to oxidation and the high Fe-$\rm L_3/L_2$ jump ratio facilitates atomic scale EELS measurement. A strong magnetic response is additionally supported by their high Curie temperatures of $\rm T_C = 740~K$, low saturation magnetization $\rm M_S = 0.38~MA/m$ and extremely high magnetic moment of $m_0 = 20~\mu_B/f.u.$ (and with $\rm Fe^{3+}$ ions up to $\rm 5~\mu_B$/atomic site) \cite{pullar2012,brabers1999}. A structural and spin model is given in Fig. 1C. The $\rm Fe^{3+}$ ions are distributed in five magnetic sublattices. Three with their magnetic spin moment parallel to the c-axis (2a, 2b and 12k) and two with antiparallel spin arrangement ($\rm 4f_1$ and $\rm 4f_2$), see fig. \ref{figS8} for details. 
Figure 1A shows the schematic diagram of the generation of isolated EVBs, the interaction with the magnetic sample and the measurement of the EEL spectra, which reflects the mapped imbalance in the local DOS for the spin up and spin down channel. Fig 1B shows the elemental mapping of the $\rm BaFe_{11}TiO_{19}$ sample in [001] direction. The $\rm Ti-L_{2,3}$ map shows a strong intensity on the $\rm 4f_2$ position in the unit cell. The expected ferrimagnetic arrangement of the magnetic moments is shown in fig. 1C. 
For the magnetic measurements, EVBs with $\boldsymbol\ell = +\hbar$ and $\boldsymbol\ell =- \hbar$ are created and scans are performed one after another. Several fast scans are averaged in order to detect the small changes in the $\rm Fe-L_{2,3}$ edge spectra \cite{Supplement}. Fig 2A shows the unit cell averaged annular dark field (ADF) images for the $\boldsymbol\ell = \pm \hbar$ EVBs. Mainly two atomic columns are visible (i) Ba-Fe column (Fe on $\rm 4f_1 $ and $\rm 4f_2$) with a high intensity in the ADF image and (ii) a pure Fe column (Fe on 2a, 2b) with a slightly weaker ADF intensity. According to the magnetic ground state of the ferrimagnet $\rm BaFe_{12}O_{19}$ \cite{collomb1986}, these two columns have antiparallel magnetic spin moments (see. Fig 1C). The acquired spectra are aligned and post-edge normalized. The $\rm Fe-L_3$ and $\rm Fe-L_2$ edge are fitted using a convolution of Lorentzian and Gaussian functions (see Supplement). The elemental mappings of the $\rm Fe-L_3$ intensity for the two OAM states are shown in fig. 2B. Note, that the 2a,b Fe columns show a higher intensity in the $\rm Fe-L_{3}$ edge map than the $\rm 4f_{1,2}$ Fe columns due to the atomic column density difference. Only in the case of Ti doping included in the ineleastic scattering simulation, the intensity difference in the $\rm Fe-L_3$ map could be reproduced, otherwise a similar intensity was found on $\rm 4f_{1,2}$ and 2a,b atomic positions.
Fig 2C shows the calculated $\rm Fe-L_3$ edge map of a 20~nm thick Ti doped (on $\rm 4f_2$ position) $\rm BaFe_{11}TiO_{19}$ sample. The intensity distribution of the 2a,b and $\rm 4f_{1,2}$ atomic columns fit extremely well the experimental maps (fig 2B).
Figure 2D presents the extracted $\rm Fe-L_{2,3}$ spectra for the $\boldsymbol\ell = \pm \hbar $ EVBs from the 2a,b Fe-column over an area of 0.85~$\Angstrom^2$. To assess the noise levels, the raw data points are included in the spectra. The resulting difference spectra, which is the EMCD signal, reveals a distinct variation at the $\rm Fe-L_3$ edge, accompanied by an opposite signal at the $\rm Fe-L_2$ edge. This clearly demonstrates that atomically sharp EVBs are capable of measuring local magnetic moments effectively.
As a cross-check, Figure 2D displays the spectra extracted from both the 2a,b and $\rm 4f_{1,2}$ sites using the same orbital angular momentum (OAM) of the electron beam. Due to the ferrimagnetic ordering, an EMCD signal is also observed in this case, further validating the measurement technique. 
Assuming the 20-nm-thick sample comprises nine unit cells along the beam direction, there are a total of 45 Fe atoms, each contributing a magnetic moment of $\rm 5~\mu_B$ aligned along the beam axis. Consequently, the vortex beam EMCD can detect a cumulative magnetic moment of approximately $\rm 225~\mu_B$.

\section*{Magnetic Imaging and Spectroscopy}
Figure 3A, B shows the $\rm Fe-L_{2,3}$ EMCD images obtained by subtracting the $\rm Fe-L_{2,3}$ edge maps (fig. 2B) for the $\boldsymbol\ell = \pm \hbar $ EVBs, respectively. Here, blue represents a negative EMCD signal, whereas red is assigned to a positive EMCD signal. 
The strongest positive EMCD signal can be found on the 2a,b Fe columns which is due to their magnetic moments oriented parallel to the electron beam. Even though, the ADF images (fig 2A), show a clear atomic structure with no hints to strong imaging aberrations, the EMCD maps seem to have residual aberrations visible by the bended intensity distributions. The EMCD signal has a sign change on the $\rm 4f_{1,2}$ column and with a reduced strength as compared to the 2a,b column. This can be interpreted as an antiparallel magnetic moment with reduced magnitude, typical for a ferrimagnet. This is in accordance with the area spectra in fig 2D, showing an EMCD signal extracted from 2a,b and $\rm 4f_{1,2}$ column  using the same OAM beam. 
Due to the intensity difference of the $\rm Fe-L_3$ and $\rm Fe-L_2$ edges in the EEL spectra, the $\rm Fe-L_3$ EMCD map shows a higher signal strength as compared to the $\rm Fe-L_2$ EMCD map and is by this more susceptible to noise. However, it is clearly visible, that the $\rm Fe-L_2$ EMCD map shows the opposite color coding and therefore reflects the same magnetic moment configuration as discussed for the $\rm Fe-L_3$ EMCD map. 
Atomically resolved EELS, and in particular EMCD, strongly depends on the dynamic diffraction conditions of the investigated sample \cite{Rusz2013, Rusz2014, song2019}. In order to verify the influence of sample thickness, collection angle and Ti doping, inelastic scattering simulations have been performed. 
The $\rm Fe-L_3$ EMCD map shown in fig. 3C resembles the experimentally observed EMCD map (fig. 3A). Further simulations including different doping levels are included in the Supplement.

\section*{Doping Effects}
A comparison between the simulated and experimental EMCD maps reveals a significant difference in the symmetry of the unit cell. The simulation exhibits the expected 6-fold symmetry characteristic of hexagonal Ba-hexaferrite, while the experimental EMCD images display a 3-fold symmetry. This discrepancy suggests potential variations in the experimental conditions or sample characteristics that may influence the observed symmetry, warranting further investigation into the underlying causes.
A significant variation in EMCD signal is observed between two adjacent 4f$_{1,2}$ columns. The signal exhibits notable weakening in one column, with the average signal across the atomic column showing a sign reversal. A linescan along the (020) direction, presented in fig. 4A, traverses through the following sequence: 2a,b column $\rightarrow$ $\rm 4f_{1,2}$ column (left) $\rightarrow$ $\rm 4f_{1,2}$ column (right) $\rightarrow$ 2a,b column. While crystal symmetry would predict a symmetric magnetic EMCD signal, the observed asymmetry shows a direct correlation with the titanium concentration distribution in the crystal structure (Figure 1B). The titanium distribution exhibits a non-homogeneous pattern across the $\rm 4f_2$ positions, preferentially forming a superstructure that results in a 3-fold crystal symmetry. This arrangement produces two significant effects: (i) The nonmagnetic $\rm Ti^{4+}$ ions reduce the total magnetic moment of the $\rm 4f_{1,2}$ column and (ii) a charge transfer occurs at the neighboring Fe atoms, leading to a reduced magnetic moment.
Detailed DFT calculations incorporating titanium doping at the $\rm 4f_2$ site reveal substantial modifications in the density of states (DOS), as illustrated in figure 4B (see fig. \ref{figS6} for comparison). The DOS reveals a characteristic shift of the spin-down channel below the Fermi energy, transforming the crystal into a half-metallic state (see inset of fig. \ref{fig4}B). The occupied spin-down states just below the Fermi energy arise from Fe 3d orbitals at the 12k and 2b sites near the Ti dopant, forming a locally conductive Fe cluster. This charge transfer from Ti to Fe at 12k and 2b reduces the local magnetic moment on these sites. The $\rm Fe^{2+}$ ions, with their reduced magnetic moment ($\rm 4~\mu_B$) compared to undoped $\rm Fe^{3+}$ ions ($\rm 5~\mu_B$), contribute to the diminished total EMCD signal observed in the affected column. Notably, the influence of titanium doping on scattering conditions was minimal due to its comparable atomic weight with iron and insufficient to explain the observed signal weakening and sign inversion (detailed analysis available in Supplementary Information \cite{Supplement}).
The DFT calculations reveal substantial modifications in the exchange integral across different magnetic sublattices \cite{Supplement}. Due to the non-magnetic Ti dopant atom, the exchange integral $J_{4f_2,j}$ at the $\rm 4f_2$ position vanish. Using the calculated exchange energies, energy minimization calculations of the magnetic unit cell, considering various spin configurations and asymmetric titanium doping demonstrate that these modifications induce spin reorientation in the neighboring 2b sites. Fig. \ref{fig4}C shows the resulting spin configuration of $\rm BaFe_{11}TiO_{19}$ with Ti substituted at the $\rm 4f_2$ sites.  

\section*{Summary and outlook}
Our research demonstrates a significant advancement in the characterization of magnetic materials through the successful implementation of electron vortex beams (EVBs) carrying orbital angular momentum (OAM) at the $\Angstrom$ngstr\"{o}m scale. Through electron energy-loss magnetic circular dichroism (EMCD) measurements, we have achieved unprecedented resolution in revealing the atomic-scale ordering of antiparallel aligned magnetic moments within the ferrimagnetic $\rm BaFe_{11}TiO_{19}$ system. This comprehensive analysis provides crucial insights into how dopant incorporation influences both the electronic structure and magnetic properties at the atomic level. Our technique's ability to achieve atomic resolution break new ground for investigating complex magnetic phenomena via magnetic spectroscopy and imaging  that were previously inaccessible. 
We see a strong benefit for the field of spintronics and interface magnetism by enabling detailed investigation of spin structures at interfaces, characterization of magnetic heterostructures, and even orbital currents \cite{Go2018} at material interfaces. The technique shows particular promise for studying recently discovered altermagnetic materials \cite{Smejkal2022}, which represent a new class of magnets. Even an extension of the technique to study dynamic magnetic processes in real-time by using ultrafast electron microscopy \cite{fang2024} is accessible. Further developments could use ptychography to reconstruct 3D magnetization and achieve single atom sensitivity. 

\clearpage

\begin{figure}[tbp]
\includegraphics[width=0.9\textwidth]{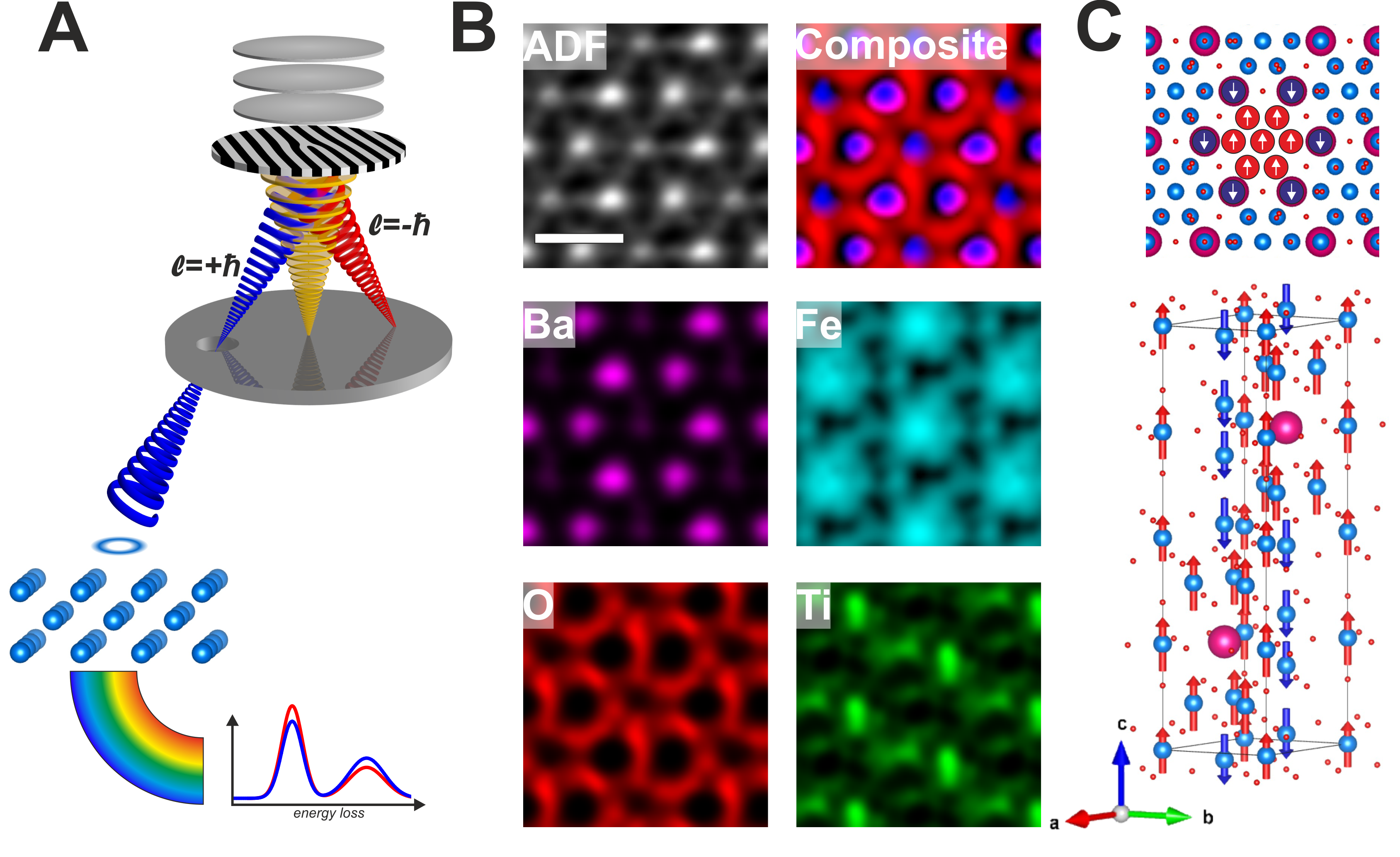}
\caption{\textbf{Generation of Vortex beams.} (A) Schematic experimental vortex EMCD setup with user-selected OAM. (B) ADF HR-STEM image and elemental mapping of $\rm BaFe_{11}TiO_{19}$ in [001] zone axis. (C) Projection of the atomic model along [001] direction and atomic spin model of the hexa-ferrimagnet $\rm BaFe_{12}O_{19}$ (magenta=Ba, red=O, blue=Fe).}
\label{fig1}
\end{figure}

\begin{figure}[tbp]
\includegraphics[width=0.5\textwidth]{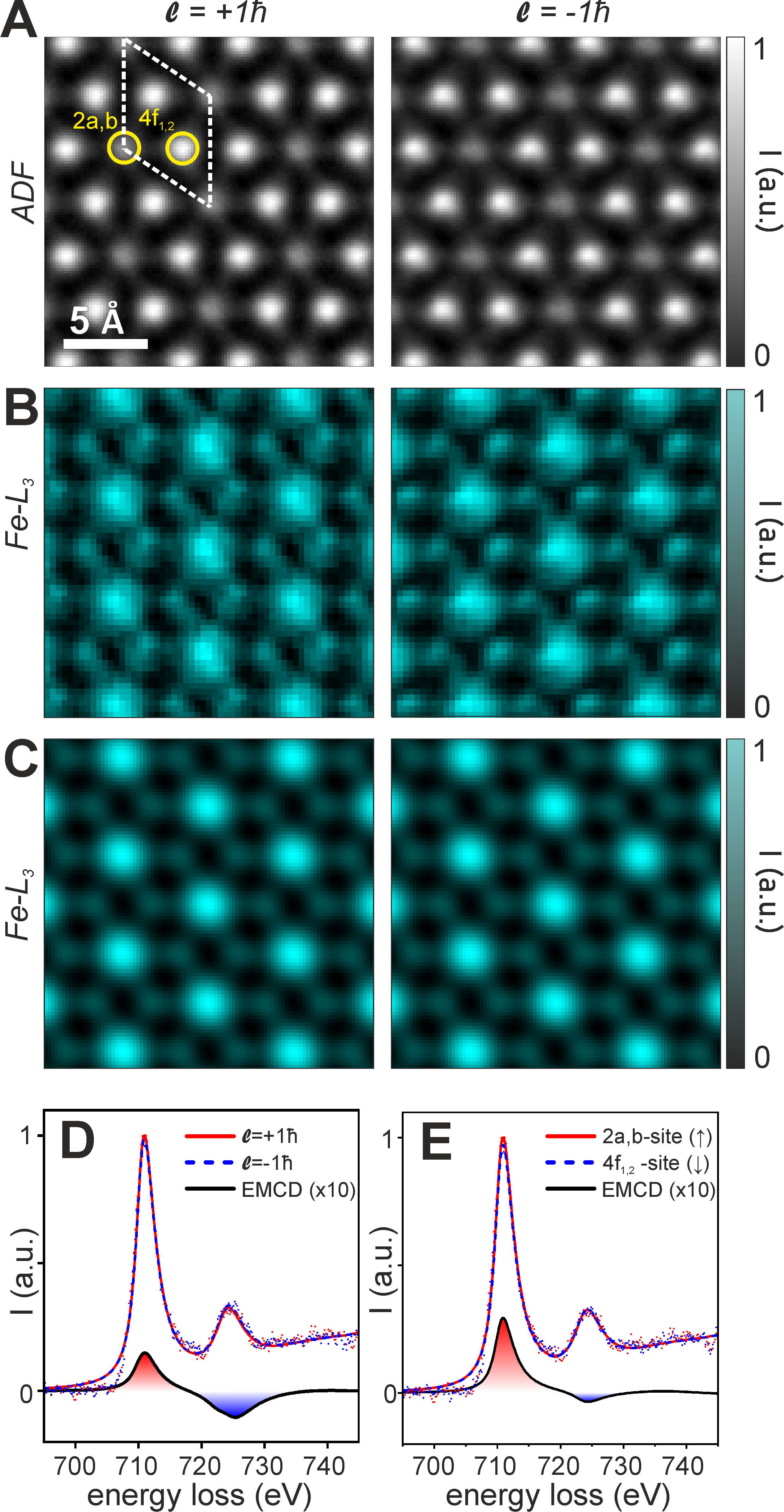}
\caption{\textbf{Vortex beams spectroscopy.} (A) Unit-cell averaged ADF image acquired with $\boldsymbol\ell = \pm \hbar $ vortex beam, respectively. The 2a,b and 4f$_{1,2}$-sites as well as the borders of the unit cell are marked by circles and dashed white lines. (B) Fe-$L_3$ image acquired with $\boldsymbol\ell = \pm \hbar $ vortex beam. (C) Inelastic scattering simulation for the $\rm Fe-L_3$ edge. (D) EEL spectra and EMCD signal extracted for the 2a,b-site atomic column. The dots represent the measured data; the lines represent the fitted data. (E) EEL spectra and EMCD extracted for the 2a,b- and $\rm 4f_{1,2}$-site (Ba-site) atomic column using $\boldsymbol\ell = + \hbar $ vortex beam. The dots represent the measured data; the lines represent the fitted data.}
\label{fig2}
\end{figure}

\begin{figure}[tbp]
\includegraphics[width=0.9\textwidth]{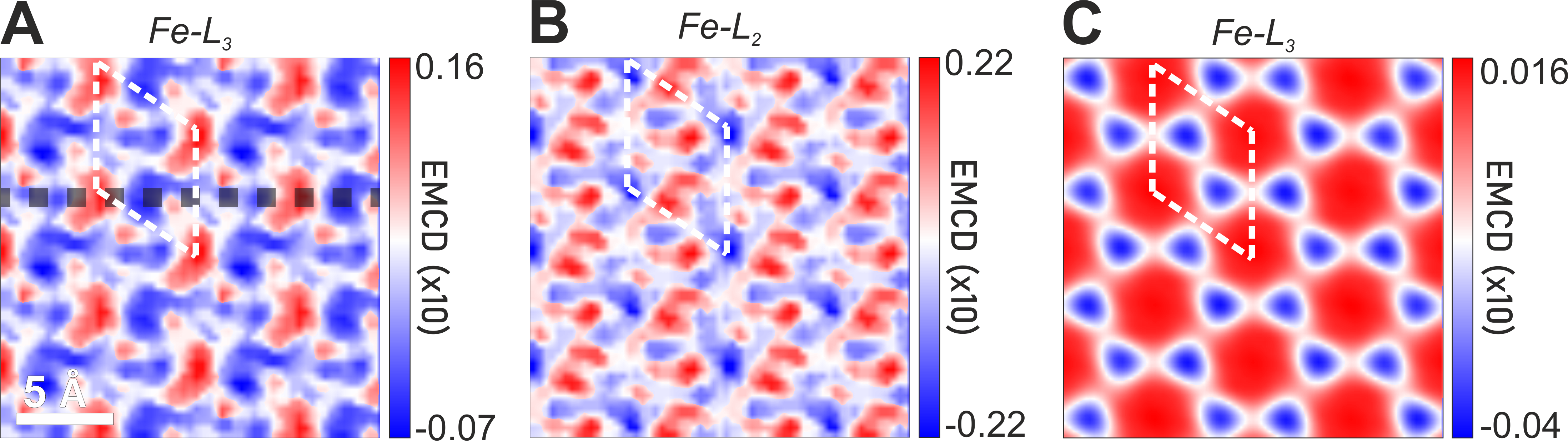}
\caption{\textbf{Magnetic spectroscopy.} (A) Experimental EMCD maps for the $\rm Fe-L_{3}$ and (B) $\rm Fe-L_{2}$ edge. The dashed lines reflect the borders of the unit cell. The thick gray line represents the intensity profile in fig. 4A. (C) Simulated EMCD map of the 20~nm $\rm BaFe_{11}TiO_{19}$ ferrimagnet.}
\label{fig3}
\end{figure}

\begin{figure}[tbp]
\includegraphics[width=0.9\textwidth]{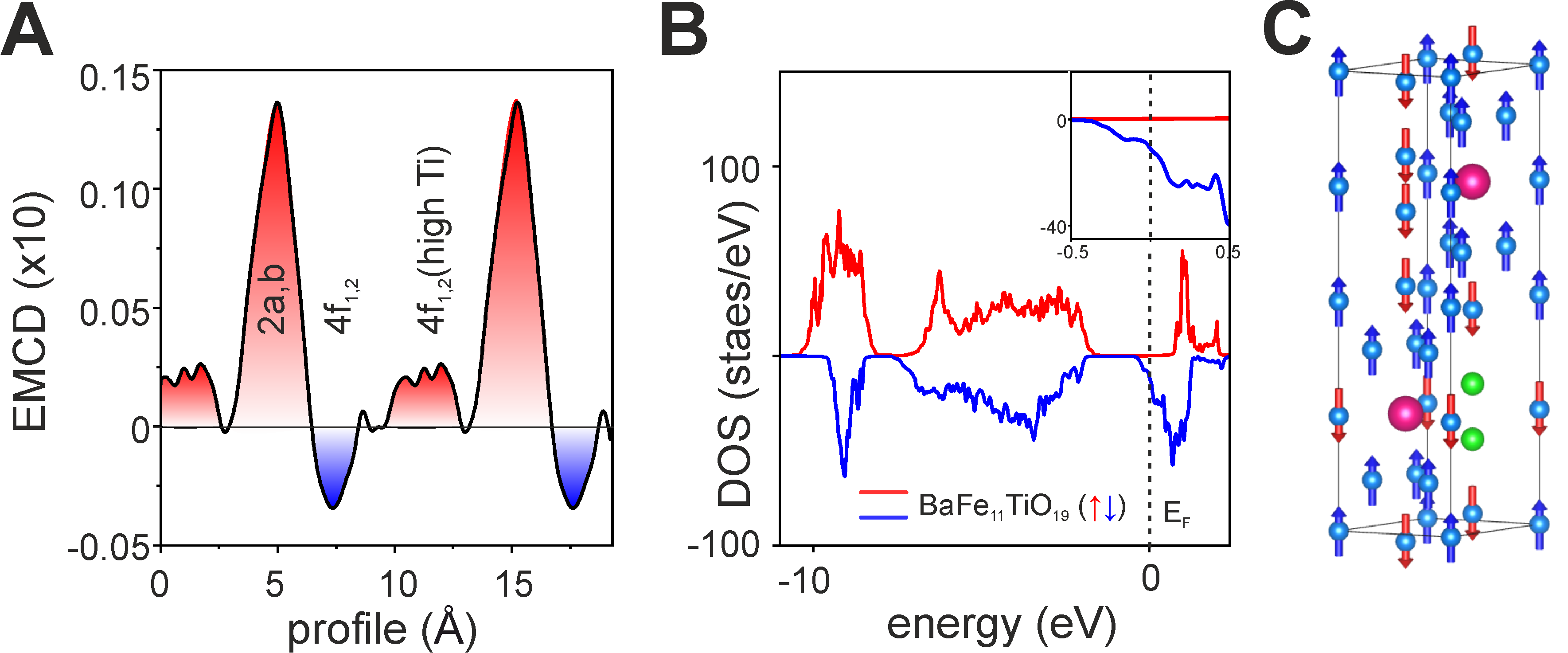}
\caption{\textbf{Local Ti doping effects} (A) EMCD intensity profile along (010), dashed line in fig 3A. (B) Spin-polarized density of states for $\rm BaFe_{11}TiO_{19}$. The inset shows a magnified view within $\rm \pm0.5~eV$ of the Fermi energy, $\rm E_F$. (C) Spin configuration with Ti (green) on $\rm 4f_2$ sites. Note the modified orientation of the magnetic moment at the 2b site, arising from altered exchange coupling near the Ti site.}
\label{fig4}
\end{figure}



\clearpage 

%
\bibliography{Pohl_et_al_2025}
\bibliographystyle{Science}

%
%
%
%
%
%


\section*{Acknowledgments}
A. E. acknowledges the funding by the Fonds zur Förderung der wissenschaftlichen Forschung (FWF) under Grant No. I 5384.
J.R.\ acknowledges the support of the Swedish Research Council (grant no.\ 2021-03848), Olle Engkvist's foundation (grant no.\ 214-0331), and Knut and Alice Wallenbergs' foundation (grant no.\ 2022.0079). The simulations were enabled by resources provided by the National Academic Infrastructure for Supercomputing in Sweden (NAISS) at PDC Centre partially funded by the Swedish Research Council through grant agreement no. 2022-06725. D.P. and H.M. are thankful for funding by the Deutsche Forschungsgemeinschaft (DFG, German Research Foundation) grant no. 504660779. We are thankful to Victor Brabers for providing the samples. We are indebted to Bernd Rellinghaus and Axel Lubk for fruitful discussions. We thank Alexander Tahn for support with sample preparation. 

\paragraph*{Funding:}
\paragraph*{Author contributions:}
D.P. designed the experiment. D.P., R.E., S.S. and H.M. performed the experiments. J.R. and D.N. did the inelastic scattering simulations. A.E. made the DFT simulations. H.M. and D.P. analyzed the experimental data. All authors contributed to writing the manuscript. 
\paragraph*{Competing interests:}
There are no competing interests to declare.
\paragraph*{Data and materials availability:}
\subsection*{Supplementary materials}
Materials and Methods\\
Supplementary Text\\
Figs. S1 to S7\\


\newpage


\renewcommand{\thefigure}{S\arabic{figure}}
\renewcommand{\thetable}{S\arabic{table}}
\renewcommand{\theequation}{S\arabic{equation}}
\renewcommand{\thepage}{S\arabic{page}}
\setcounter{figure}{0}
\setcounter{table}{0}
\setcounter{equation}{0}
\setcounter{page}{1} 


\begin{center}
\section*{Supplementary Materials for\\ \scititle}


Darius Pohl$^{1\ast\dagger}$,
Hitoshi Makino$^{1}$,
Arthur Ernst$^{2,3,4}$,
Devendra Singh Negi$^{5}$,
Sebastian Schneider$^{1}$,
Rolf Erni$^{6}$,
Jan Rusz$^{7}$\\	

\small$^\ast$Corresponding author. Email: darius.pohl@tu-dresden.de\\

\end{center}

\subsubsection*{This PDF file includes:}
Materials and Methods\\
Supplementary Text\\
Figures S1 to S8\\


\newpage


\subsection*{Materials and Methods}

\subsubsection*{Sample preparation}
Single crystals of $\rm BaFe_{11}TiO_{19}$ are grown by a floating zone technique. Details of the preparation and the magnetic measurements can be found in \cite{brabers1999}. For STEM investigation, thin electron transparent lamellae are prepared by focused ion beam technique using a FEI Helios 600i operated at 30~kV. A final polishing with 900~eV Ar ions was performed in order to remove amorphous material from the lamella surface.  

\subsubsection*{Transmission electron microscopy : Themis Thermo Fisher 300kV – data acquisition}
A Titan Themis G2 60-300, operated at 300~kV, with a XFEG and a DCOR+ corrector for correcting aberrations up to D4 is used for the here presented experiments. The corrector is tuned within the special condenser setting in order to reach atomic resolution with $\rm C_S < 1~ \mu m$. This leads to extremely fine focused beams with a convergence semi-angle of 27.7~mrad, leading to a mainly diffraction limited probe size of roughly 0.1~nm. ADF HRSTEM images are acquired with a collection angle from 59-136~mrad. For the EEL spectra acquisition, a CEOS CEFID spectrometer equipped with a DECTRIS ELA direct electron camera is used (dispersion of 0.2 eV/pixel) with a collection semi-angle of 21.8~mrad. The acquisition time of a single spectra was 0.05~s. 

\subsubsection*{Data treatment}
The unit cell averaging procedure is schematically described in fig. S1. First, the local unit cell is estimated by the Fe column, which is recognized by cross-correlation with an arbitrary kernel in the ADF image. Second, the so obtained unit cell is applied to the corresponding 3D spectrum image (SI) data, and the SI of each unit cell is extracted using the obtained boundaries. ADF image strains originating from probe scan error and spatial drift of the sample stage are corrected using an affine transformation based on the Fe column position on the unit cell corners. Background subtraction has been performed using the PCA method described in \cite{spiegelberg2017}. After averaging the ADF image and EEL spectra, the unit cell is periodically extended for clearer visualization and further analysis.
To denoise the EEL spectra and map the EMCD signal distribution, we construct a fit model consisting of Gaussian and Lorentzian functions and apply it to the raw background subtracted SI data. In advance of fitting, 2x2 pixel binning, corresponding to $0.306~\Angstrom$ x $0.306~\Angstrom$ = $0.094~\Angstrom^2$, and a Wiener smoothing filter with 3x3 pixels ($0.92~\Angstrom$ x $0.92~\Angstrom$ = $0.85~\Angstrom^2$) is applied to the averaged SI to minimize the possibility of fitting failure. The fit model is a combination of two halves of Lorentzian function (Lorentzian multiplied with a step function), which represent the $\rm Fe-L_3$ and $\rm L_2$ white lines, as well as their Fermi and tunneling tails and a Gaussian function, which represents the post-edge, including the extended energy-loss fine structure (EXELFS). The fit model $f$ is described numerically as follows:
\begin{equation}
\begin{split}
f(h_1,w_{11},w_{12},h_2,w_{21},w_{22},h_3,w_{3}, E) \\
= \{ h_1 \textbf{L}(w_{11}, E-E_{L3})\cdot \Theta(E-E_{L3} \} \otimes \textbf{G}(w_{12},E') \\
+ \{ h_2 \textbf{L}(w_{21}, E-E_{L2})\cdot \Theta(E-E_{L2} \} \otimes \textbf{G}(w_{22},E') \} \\
+h_3\textbf{G}(w_{3},E-E_{post})
\end{split}
\end{equation}
where $\textbf{L}(w,E)$ is a Lorentzian function with width $w$, $\textbf{G}(w,E)$ is a Gaussian function, and $\Theta(E)$ is a step function. The peak energies of $\rm Fe-L_3$, $\rm Fe-L_2$, and post edge are $\rm E_{L_3}=709~eV$, $\rm E_{L_2}=722~eV$, and $E_{post}=750~eV$, respectively. The experimental EELS and fitting results are shown in fig. \ref{figS2}(A). We assumed that the widths of each function, $w$, do not change spatially. Therefore, we used the median values of the free fitting over the entire SI data set and fixed the widths during the subsequent fit. Thus, spatial distributions of the coefficient of $h_1$, $h_2$, and $h_3$ represent the signal intensities mappings of the $\rm Fe-L_3$, $\rm Fe-L_2$, and post-edge, respectively, as shown in Fig. 2(B) and \ref{figS2}(B). 
In order to validate the accuracy of the fitting procedure, the residual maps for the $\rm Fe-L_{2,3}$ edges are shown in fig. S5 together with the fit coefficients $h_1$, $h_2$, and $h_3$. 

To evaluate the EMCD signal, the fitted spectra are normalized in the energy region of 717.5-718.5~eV. This energy region is chosen, to reduce the effect of the background subtraction and is known from own classical EMCD experiments to show no EMCD signal.  
The EMCD maps shown in Fig. 3(A) are simply calculated as:
\begin{equation}
    EMCD_{L3}=\frac{h_{1,norm}(\boldsymbol\ell = + \hbar)-h_{1,norm}(\boldsymbol\ell = - \hbar)}{h_{1,norm}(\boldsymbol\ell = + \hbar)+h_{1,norm}(\boldsymbol\ell = - \hbar)},
\end{equation}
and
\begin{equation}
     EMCD_{L2}=\frac{h_{2,norm}(\boldsymbol\ell = + \hbar)-h_{2,norm}(\boldsymbol\ell = - \hbar)}{h_{2,norm}(\boldsymbol\ell = + \hbar)+h_{2,norm}(\boldsymbol\ell = - \hbar)},
\end{equation}
where $\rm EMCD_{L_3}$ and $\rm EMCD_{L_2}$ are the EMCD intensities of $\rm Fe-L_3$ and $\rm Fe-L_2$ edges, respectively. To reduce noise, a 3x3 pixel median filter is applied. 

In total, to achieve the necessary signal-to-noise ratio (SNR) for effective mapping, averaging was performed over 147 unit cells for $\boldsymbol\ell = + \hbar$ and 198 unit cells for $\boldsymbol\ell = - \hbar$. Given that each unit cell occupies an area of $30.4~\Angstrom^2$, this results in an averaged area of roughly $44.7$~nm$^2$ for $\boldsymbol\ell = + \hbar$ and $60.2$~nm$^2$ for $\boldsymbol\ell = - \hbar$. This averaging process enhances the reliability of the EMCD measurements, allowing for more accurate mapping of the magnetic properties within the sample. Still, the calculation of $m_L/m_S$ maps is beyond the current SNR level. 

\begin{figure}[tbp]
\includegraphics[width=0.8\textwidth]{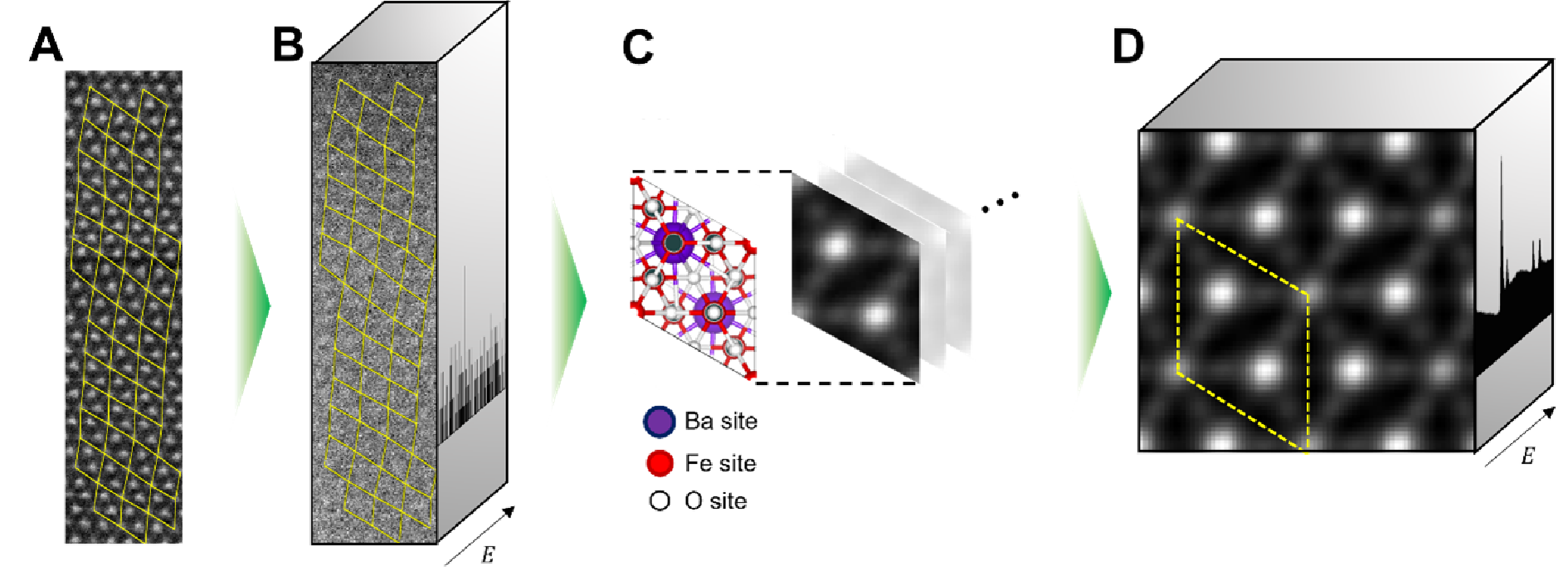}
\caption{\textbf{Unit-cell averaging.} (A) Local unit cell definition based on atomic column positions in the ADF image. (B) Extraction of the unit cells from the raw SI datasets. (C) De-strain the unit-cells by affine transformation and averaging. (D) Periodic extension of the averaged unit cell for visualization.}
\label{figS1}
\end{figure}

\begin{figure}[tbp]
\includegraphics[width=0.8\textwidth]{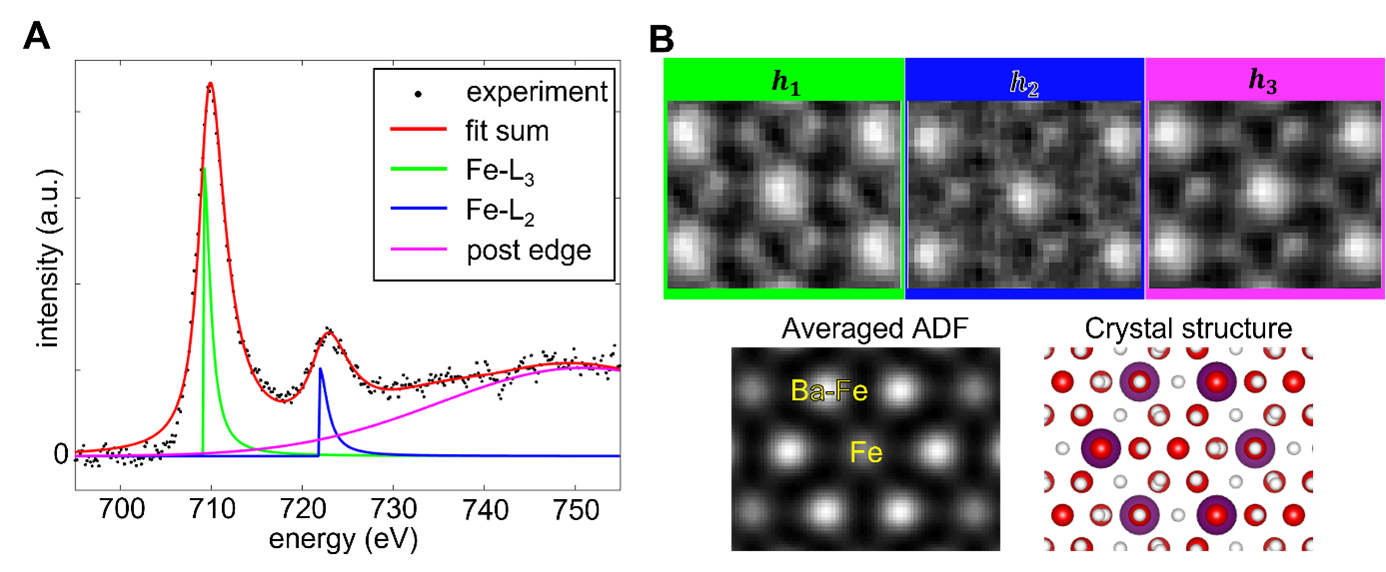}
\caption{\textbf{Denoising by Gaussian and Lorentzian fitting} (A) Exemplary experimental EELS data and the resulting fit including the single fit functions at $\rm Fe-L_3$, $\rm Fe-L_2$ and post edge region. (B) Coefficient maps for the parameters $h_1$, $h_2$, and $h_3$. Lower panel shows averaged ADF image and the expected crystal structure. }
\label{figS2}
\end{figure}

\begin{figure}[tbp]
\includegraphics[width=0.8\textwidth]{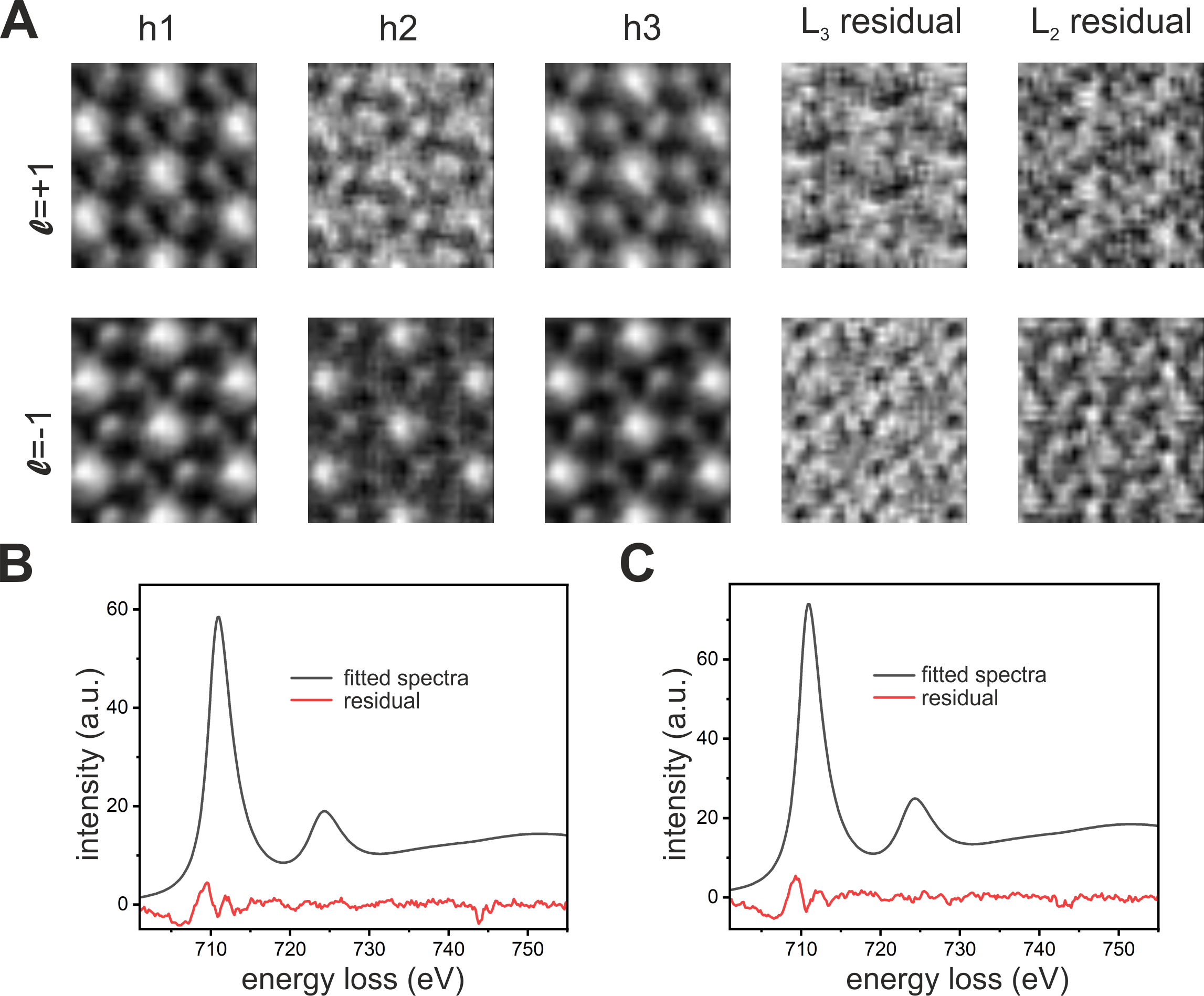}
\caption{\textbf{Residual of fitted spectrum images} (A) Fit coefficient maps and residuals of the fit for the $\boldsymbol\ell = \pm \hbar$ spectrum images. The residual maps for the $\rm Fe-L_{2,3}$ edges are integrated from 710.6-711.4~eV and 723.9-724.7~eV, respectively. (B) Fitted EEL spectra (black solid line) and residual (red solid line) for the 2a,b central atomic column integrated over 5x5 pixels of the $\boldsymbol\ell = + \hbar$ spectrum image. (C) Fitted EEL spectra (black solid line) and residual (red solid line) for the 2a,b central atomic column integrated over 5x5 pixel of the $\boldsymbol\ell = - \hbar$ spectrum image. }
\label{figS5}
\end{figure}

\subsubsection*{Inelastic scattering simulations}
$\rm BaFe_{12}O_{19}$ calculations were performed on an orthogonal supercell of lateral size 3.1~nm x 3.0~nm. Real space sampling in the multislice calculations used pixels of dimension 5.4~pm x 5.3~pm. Sample thicknesses of up to 30.4~nm were calculated. Acceleration voltage and convergence semi-angle were set according to experiment to 300~kV and 27.7~mrad, respectively. For inelastic scattering calculations we have used combined multislice/Bloch-waves method implemented in MATS.v2 software \cite{rusz2017mats}. 

Diffraction patterns are calculated in a range of 35~mrad in both scattering directions. The summation convergence parameter was set to $10^{-5}$. Effects of Ti doping on the inelastic scattering were modelled by subtracting the Fe atomic contributions from the substituted atomic sites.
The EMCD maps are calculated by the relative signal of magnetic to non-magnetic scattering maps \cite{negi2019}.  
In order to account for aberrations and a finite source size, the simulated inelastic scattering maps are blurred using a Gaussian with a standard deviation of $\rm 2~\Angstrom$.

\subsubsection*{DFT calculations}
First-principles calculations of electronic and  magnetic properties of pristine and Ti doped $\rm BaFe_{12}O_{19}$ were performed within the density functional theory using a generalized gradient approximation (GGA)~\cite{Perdew1996}. Strongly localized Fe 3$d$ electrons were treated within a GGA+U method~\cite{Anisimov1991,Dudarev1998}. The effective parameter $U^*=U-J=6$~eV was chosen in such a manner to obtain correct Curie temperature for the pristine  $\rm BaFe_{12}O_{19}$, estimated within a random-phase approximation (T$^{RPA}_C$=838~K vs. T$^{exp}_C$=731~K). The same effective parameter $U^*$ was used for both pristine and Ti-doped  $\rm BaFe_{12}O_{19}$. Electronic structure calculations were performed using a self-consistent full potential Green function method implemented  within the multiple scattering theory~\cite{Geilhufe2015,Hoffmann2020}. The Green function was expressed via an angular momentum expansion up to $l_{max}$=3. The crystalline structure was adopted from the current experiment. 

The Heisenberg exchange parameters $J_{ij}$ were calculated using the magnetic force theorem as it is implemented  within the multiple scattering theory~\cite{LKAG1987}. Curie temperature was estimated with a random phase approximation (RPA)~\cite{Tyablikov1995}.  

\begin{table}[ht]
\centering
\begin{tabular}{lccccc}
\hline
Element & Position & sublattice & x & y & z \\
\hline
Fe & 1 & 2a & 0.0000000 & 0.0000000 & 0.0000000 \\
Fe & 2 & 2a & 0.0000000 & 0.0000000 & 11.5494999 \\
Fe & 3 & 2b & 0.0000000 & 0.0000000 & 5.7747500 \\
Fe & 4 & 2b& 0.0000000 & 0.0000000 & 17.3242500 \\
Fe & 5 & 4f1 & 3.3861594 & 0.0000000 & 22.4522280 \\
Fe & 6 & 4f1 & 1.6930797 & 2.9325001 & 0.6467720 \\
Fe & 7 & 4f1 & 1.6930797 & 2.9325001 & 10.9027280 \\
Fe & 8 & 4f1 & 3.3861594 & 0.0000000 & 12.1962720 \\
Fe & 9 & 4f2 & 1.6930797 & 2.9325001 & 4.3818803 \\
Fe & 10 & 4f2 & 3.3861594 & 0.0000000 & 18.7171196 \\
Fe & 11 & 4f2 & 1.6930797 & 2.9325001 & 7.1676197 \\
Fe & 12 & 4f2 & 3.3861594 & 0.0000000 & 15.9313802 \\
Fe & 13 & 12k & 1.6930797 & 0.0000000 & 2.5016217 \\
Fe & 14 & 12k & 3.3861594 & 2.9325001 & 20.5973783 \\
Fe & 15 & 12k & 4.2326992 & -1.4662500 & 2.5016217 \\
Fe & 16 & 12k & 0.8465398 & 1.4662500 & 20.5973783 \\
Fe & 17 & 12k & 4.2326992 & 1.4662500 & 2.5016217 \\
Fe & 18 & 12k & 0.8465398 & 4.3987501 & 20.5973783 \\
Fe & 19 & 12k & 1.6930797 & 0.0000000 & 9.0478783 \\
Fe & 20 & 12k & 4.2326992 & -1.4662500 & 9.0478783 \\
Fe & 21 & 12k & 4.2326992 & 1.4662500 & 9.0478783 \\
Fe & 22 & 12k & 3.3861594 & 2.9325001 & 14.0511217 \\
Fe & 23 & 12k & 0.8465398 & 1.4662500 & 14.0511217 \\
Fe & 24 & 12k & 0.8465398 & 4.3987501 & 14.0511217 \\
\hline
\end{tabular}
\caption{Coordinates (in $\Angstrom$) of the 24 Fe atomic sites in $\rm BaFe_{12}O_{19}$ used for DFT calculations.}
\label{tableS1}
\end{table}

\begin{figure}[tbp]
\includegraphics[width=0.5\textwidth]{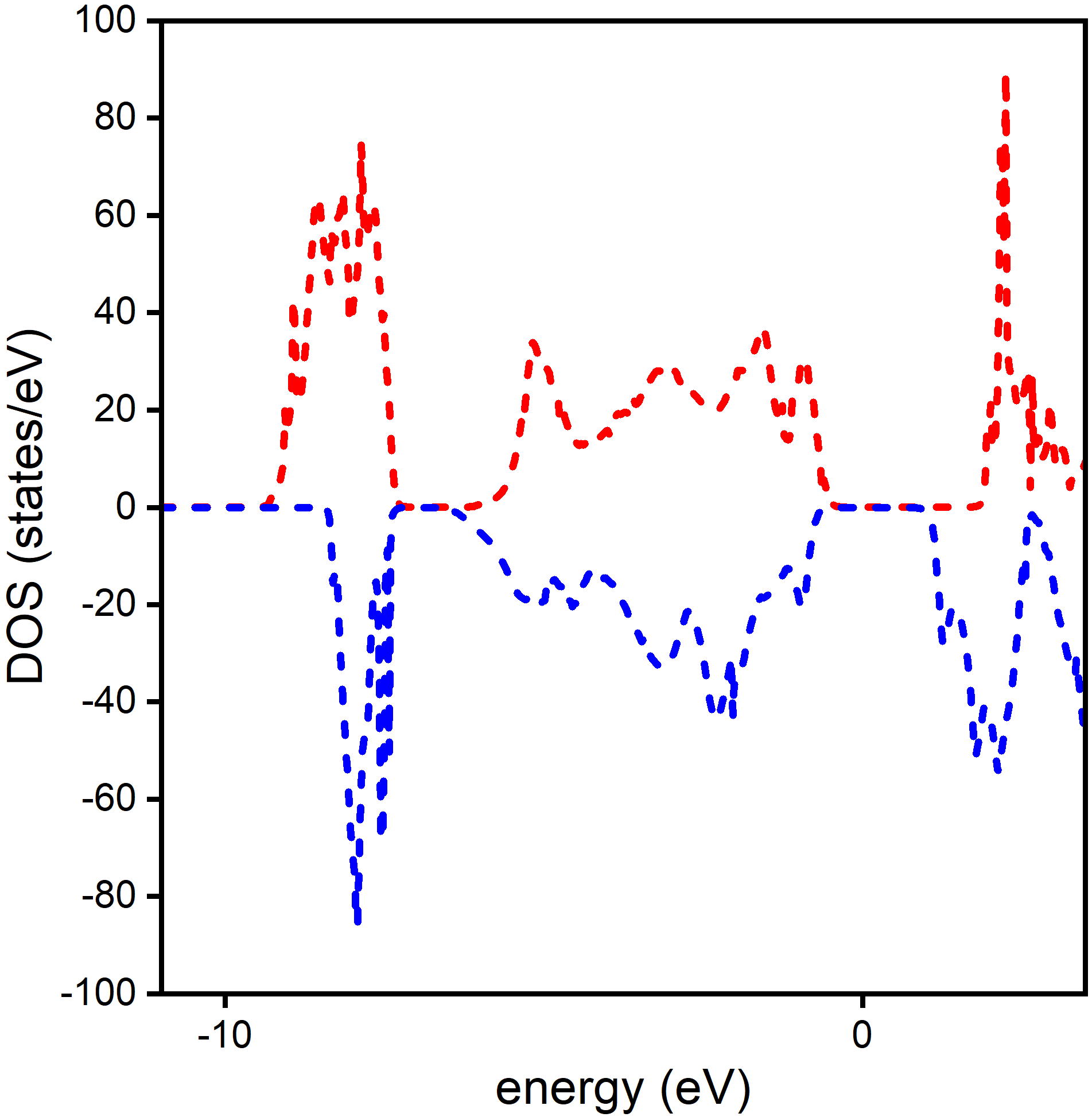}
\caption{\textbf{Spin-polarized density of states of the undoped $\rm BaFe_{12}O_{19}$. The blue dashed line represents the spin-down DOS, and the red dashed line represents the spin-up DOS. Note that the Fermi energy is in the gap, which is typical for an insulator.}}
\label{figS6}
\end{figure}

\begin{figure}[tbp]
\includegraphics[width=0.8\textwidth]{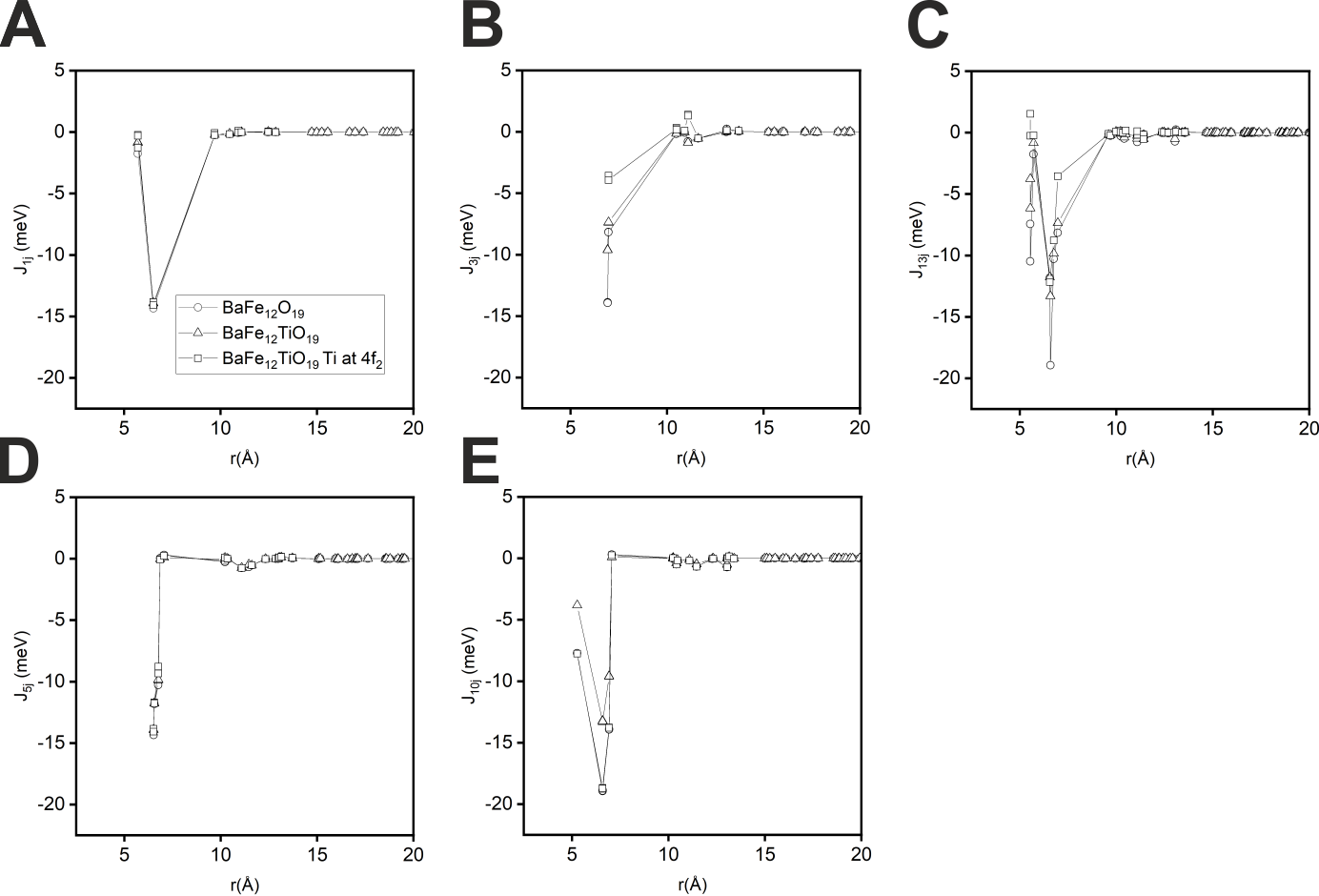}
\caption{\textbf{Exchange energies $J_{ij}$ for the different atomic sites for  $\rm BaFe_{12}O_{19}$,  $\rm BaFe_{11}TiO_{19}$, and  $\rm BaFe_{11}TiO_{19}$ with Ti only at position 9 and 11. See table \ref{tableS1} for details.} (A-E) $J_{ij}$ for the atomic position 1,3,5,10, and 13. All other position of the 24 Fe position show same behavior due to symmetry.}
\label{figS7}
\end{figure}

\begin{figure}[tbp]
\includegraphics[width=0.8\textwidth]{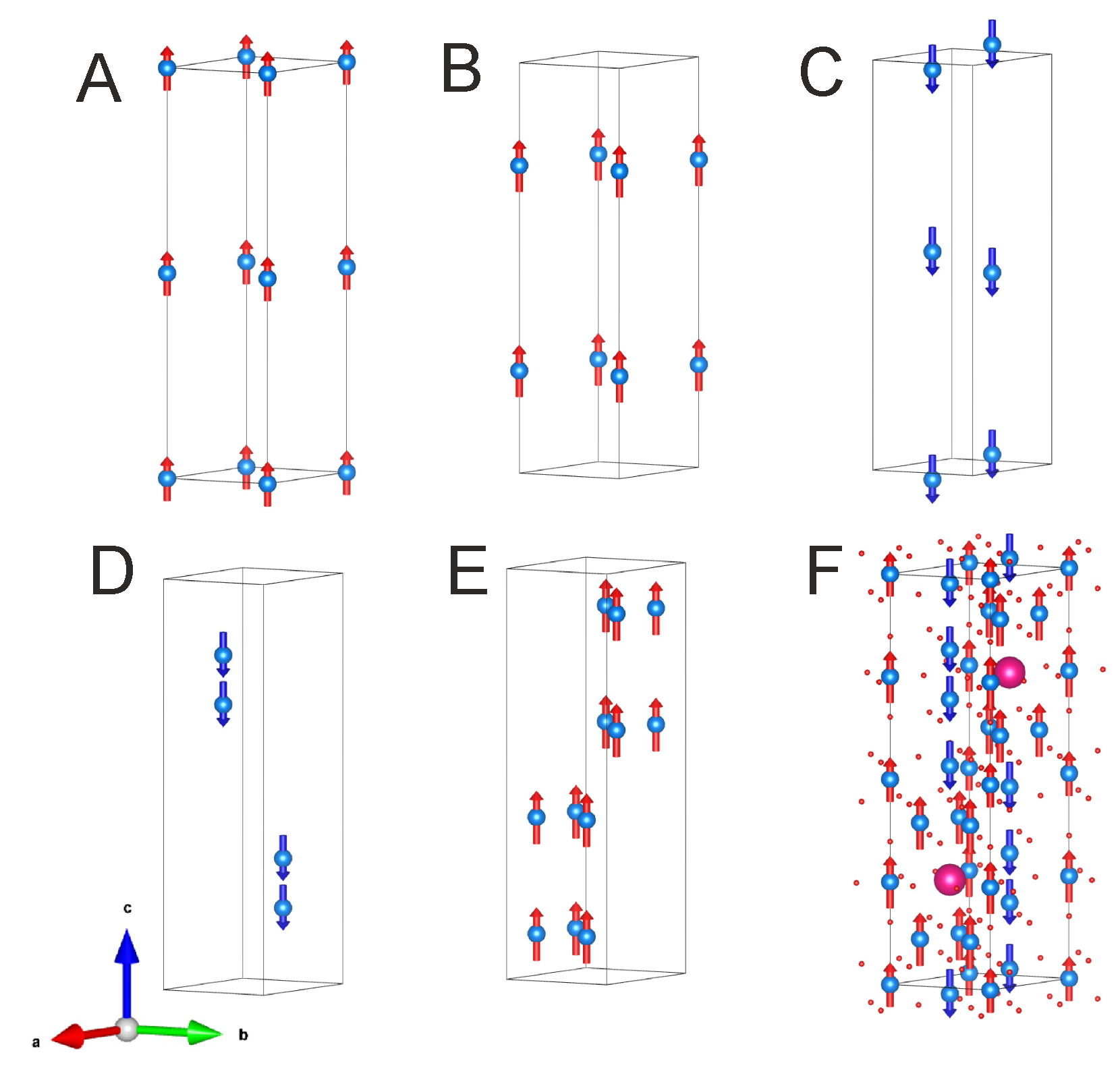}
\caption{\textbf{Unit cell representation of the five different magnetic Fe sublattices of $\rm BaFe_{12}O_{19}$.} Following the Wyckoff nomenclature the five sublattices are 2a (A), 2b (B), $\rm 4f_1$ (C), $\rm 4f_2$ (D) and 12k (E). (F) represents the unit cell with all Fe atomic positions.}
\label{figS8}
\end{figure}

In order to find the lowest energy configuration with respect to the spin orientation, a supercell of 3x3x3 unit cells of  $\rm BaFe_{12}O_{19}$ was generated. Using the exchange interactions $J_{ij}$ given by DFT simulations, the total energy for all possible spin configurations have been calculated. Hereby, the symmetry of the magnetic unit cells haven been considered in order to reduce computation effort \cite{Novak2005}. In the undoped system, the ground state with the lowest energy is given by the expected ferrimagnetic order: 2a(up), 2b(up), 4f$_1$(down), 4f$_2$(down), 12k(up). In order to evaluate the effect of doping, the symmetry has been reduced to up to nine magnetic sublattices.   


\subsection*{Supplementary Text}

\subsubsection*{Ineleastic scattering simulation - Ti doping}

\begin{figure}[tbp]
\includegraphics[width=0.5\textwidth]{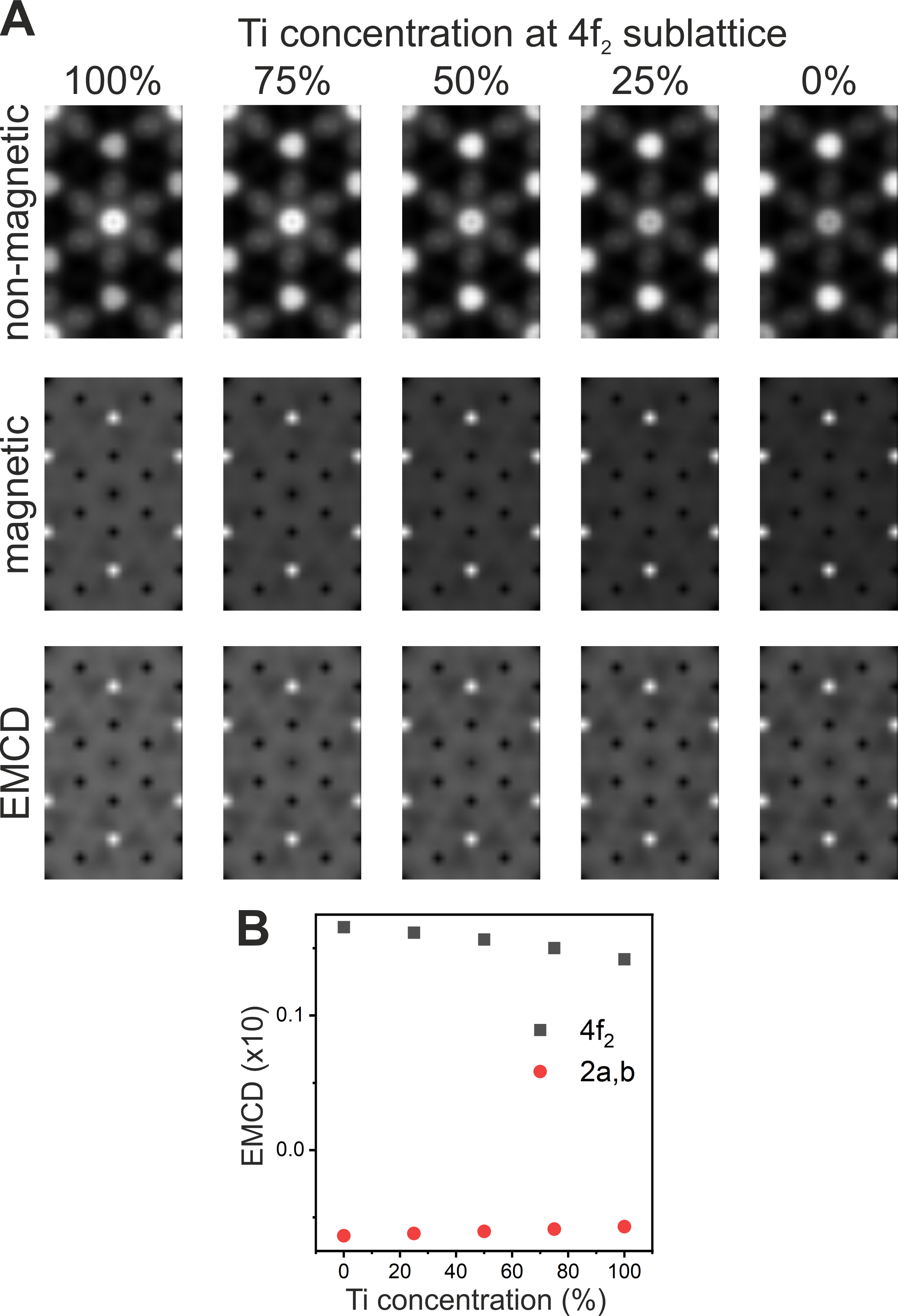}
\caption{\textbf{Simulation of the influence of Ti doping on the EMCD signal.} (A) Non-magnetic, magnetic and calculated relative EMCD maps of the 20~nm thick $\rm BaFe_{12-x}Ti_xO_{19}$ crystal along [001] direction. (B) Extracted EMCD signal for the $\rm 4f_2$ and 2a,b atomic column position, averaged over 5x5 pixels.}
\label{figS3}
\end{figure}

As described in the methods section, the effects of Ti doping on the inelastic scattering were modelled by subtracting the Fe atomic contributions from the substituted atomic sites.
Fig. \ref{figS3} summarizes the outcome of the inelastic scattering simulation, including Ti doping on the $\rm 4f_2$ sub-lattice. Only a slight variation of the intensity distribution is visible in the images. For a quantification of the influence of the doping on the EMCD signal strength, the EMCD signal is extracted and averaged over a 5x5 pixel area and plotted in fig. \ref{figS3}(B). No sign change or strong variation of the EMCD signal is observed, indicating that the influence of Ti doping on the magnetism extends beyond the simple substitution of Fe atoms.  

\subsubsection*{Thickness measurements and data selection}

The thickness of the measured area was determined by using low-loss EELS. A relative thickness of $t/\lambda = 0.18$ was measured, which corresponds to an absolute thickness of 20~nm in our case. Since no simultaneous low-loss and high-loss EELS were possible during data acquisition, we used the $\rm L_3/(post-edge)$ intensity as a measure of the thickness variation of the individual SI. Only datasets with an $\rm L_3/(post-edge)$ intensity variation less than 10\% of the mean value have been chosen for further data processing.

\begin{figure}[tbp]
\includegraphics[width=0.5\textwidth]{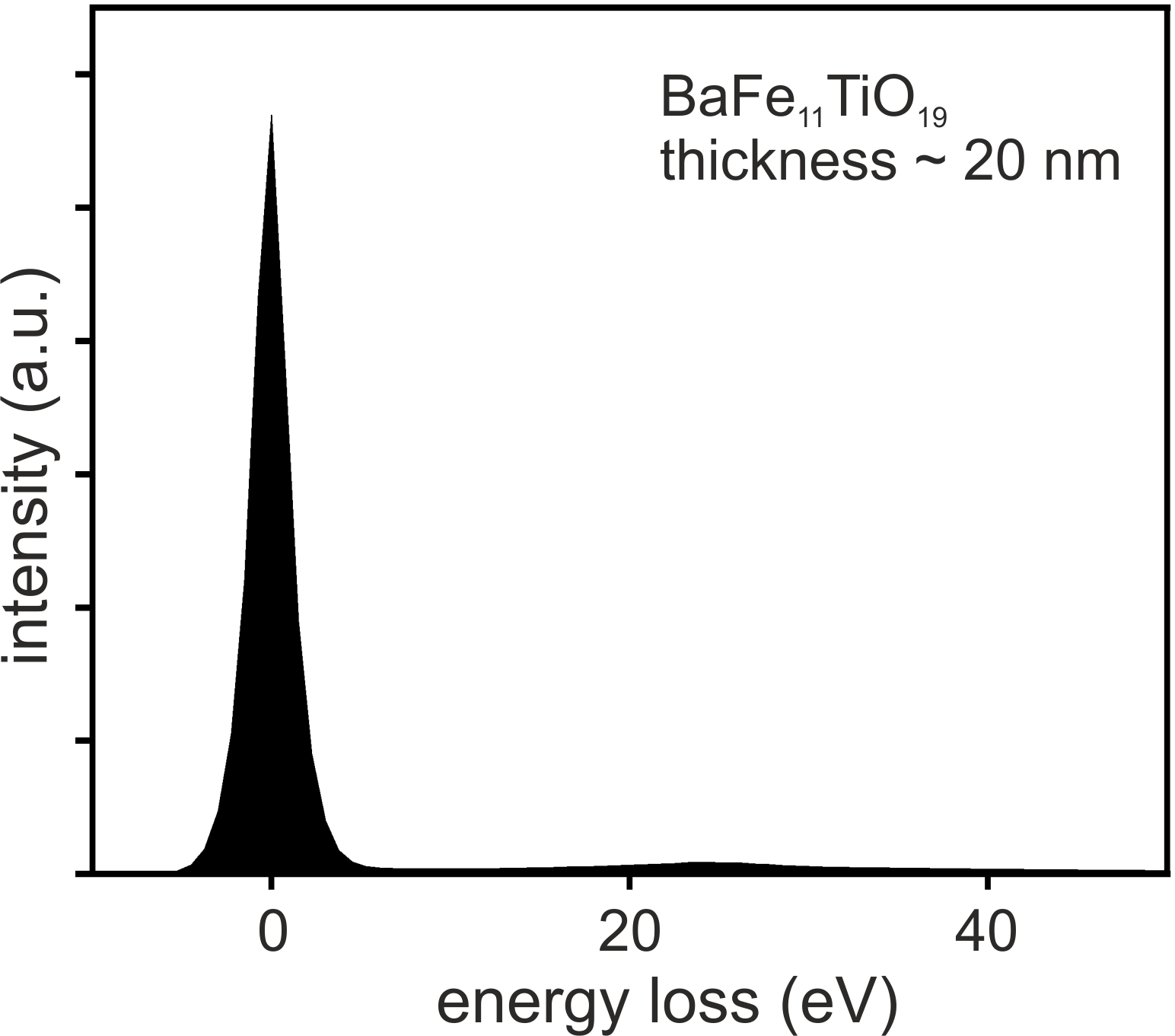}
\caption{\textbf{Thickness measurement by low-loss EELS.} Low-Loss EEL spectra obtained from the measured region on the $\rm BaFe_{11}TiO_{19}$ sample. A thickness of 20~nm determined.}
\label{figS4}
\end{figure}




\clearpage 



\end{document}